\documentclass
{aa}
\usepackage{fleqn}

\usepackage{url}

\usepackage{graphicx}
\usepackage[figuresright]{rotating}
\usepackage{natbib}
\bibpunct{(}{)}{;}{a}{}{,}
\begin{document}

\title{Astrophysical significance of detection of coalescing binaries with gravitational waves}
\titlerunning{Astrophysical significance...}
\authorrunning{T. Bulik, K. Belczy\'nski and  B. Rudak}
\author{T. Bulik \inst{1},
       K. Belczy\'nski \inst{2,3},
        B. Rudak \inst{1}
       }

\institute{$^1$Nicolaus
Copernicus Astronomical Center, Bartycka 18, 00716 Warsaw, Poland\\
          $^2$ Northwestern University, 2145 Sheridan Rd. Evanston, IL, USA\\
	  $^3$ Lindheimer Postdoctoral Fellow}


\date{Received / Accepted }

\abstract{
We use the StarTrack stellar population synthesis code to analyze properties of
double compact object binaries as sources of gravitational waves.
Since the distribution of lifetimes of  these objects
extends up to the Hubble time we conclude that a proper calculation of the
expected rate must include a full cosmological model. We present such model,
calculate the expected coalescence rates, and analyze
the intrinsic sensitivity of these
rates to the
model assumptions. We find that the rate alone is a very poor
estimator of the underlying stellar evolution model. However we show that the
distribution of observed chirp masses is very sensitive to the underlying
stellar evolution model, while it is very insensitive to the underlying
cosmology, star formation rate history and variation of
 detector sensitivity.
}

\maketitle

\section{Introduction}

The large interferometric gravitational wave detectors are entering the realm 
of astronomy: LIGO has now completed it first two science runs, 
and is improving 
its sensitivity, VIRGO preparations are proceeding very well.
GEO~600 has achieved an astonishing duty cycle above 97\%,
and TAMA~300 has been taking data in coincidence with LIGO.
The sensitivity of all the instruments is improving and 
this leads to increased hopes for detecting gravitational waves.
While a detection of gravitational wave will be a huge triumph by 
itself,
the following  question appear: can there be any useful
astrophysics done with gravitational waves? 
can gravitational wave observations provide any constraints on
astrophysical models, and be used to answer questions related to
what is known as standard astronomy?

Coalescing compact object
binaries  are one of the most promising, if not
the most promising, sources of gravitational waves. First, we know
from electromagnetic observations that they exist and that they
emit gravitational waves. Moreover, the known sources will
coalesce. The research on the properties of coalescing compact
object binaries has concentrated so far on calculation of the
expected rates. This problem has been approached in two ways.
The first approach was based on analyzing the properties of the
existing systems, and considering all the possible selection
effects that may affect detectability of such systems
\citep{1991ApJ...379L..17N,2001ApJ...556..340K}. The drawback of
this approach is the small number statistics, or even  zero
object statistics, in the case of black hole neutron star or
double black hole binaries. Moreover, the estimates of different
selection effects carry additional uncertainty.  A second
approach is based on studying the stellar evolution processes
and detailed analysis of the formation paths of double compact
object binaries
\citep{1997MNRAS.288..245L,1998ApJ...496..333F,1998A&A...332..173P,1998ApJ...506..780B,1999ApJ...526..152F,1999A&A...346...91B,1999MNRAS.309..629B,2002ApJ...572..407B,2003Nutz}.
Within this approach the main problem is insufficient knowledge
of certain processes in the stellar evolution. Such processes 
were parameterized and the results appear to vary significantly
with some of this parameters. The uncertainty on the 
expected rate of compact object coalescences is between
two and three orders of magnitude.

In our previous paper \citep{2003ApJ...589L..37B} we pointed out another
possibility: we have shown that measurement of chirp masses
of coalescing systems carries a lot of information about the 
underlying stellar evolution. This calculation
used a set of simplifying assumptions: Euclidean space geometry
and a constant star formation rate.  In this paper we relax
these assumptions and consider a more general model. In section 2
we describe the population synthesis model and argue for the need
to consider a fully cosmological model of the source distribution
to  calculate the distributions of observable quantities.
Section 3 contains an outline of such calculations, and a
discussion of the dependence of the final results on various 
cosmological parameters and assumptions. Finally we finish we
conclusions in Section 4

\section{Population synthesis implications}

In this work we use the Star Track population synthesis  code
\citep{2002ApJ...572..407B}. The single star evolution in the
code is based on the approximate analytical formulas compiled by
\citet{2000MNRAS.315..543H}. The single star evolution
description  includes such stages as the main sequence,
Hertzsprung gap evolution, red giant branch,  core helium
burning, asymptotic giant branch, and the helium star evolution.
The end product of the stellar evolution can be  a white dwarf,
a neutron star, or a black hole.  The binary evolution
description includes such  processes as variation of the orbits
due to wind mass loss, tides, as well as magnetic breaking, and
various mass transfer modes: conservative, quasi dynamic, common
envelope (CE) evolution. We also take into account the variation of
the structure of a star in response to accretion (e.g.
rejuvenation) and a possibility  of hypercritical accretion onto
compact objects. Supernovae explosions are treated
in detail using the results  of hydrodynamical simulations
\citep{1999ApJ...522..413F}. We take into account both direct 
and    fall back black hole formation. Finally, we also
investigate various kick velocity distributions. The initial
mass of the primary is drawn from a power law distribution
$\propto M^{-2.7}$ \citep{1986FCPh...11....1S}, the initial 
mass ratio distribution is flat \citep{1935PASP...47...15K},
 the   eccentricities are drawn
from a distribution   $\propto e$ \citep{1975MNRAS.173..729H,1991A&A...248..485D} , and the orbital separation
distribution is flat in $\log a$ \citep{1983ARA&A..21..343A}.   
 
Apart from the standard model (model A) described in detail by
\citet{2002ApJ...572..407B} we  also consider a variety
of different population synthesis models where we vary
parameterizations of  stellar evolutionary stages. A list
of models used later on is presented in
Table~1.

\begin{figure}
\includegraphics[width=0.9\columnwidth]{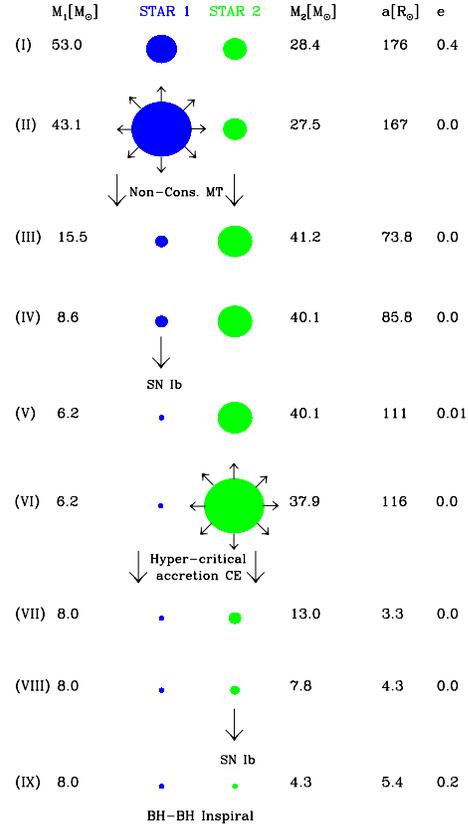}
\caption{An example evolutionary scenario leading to formation of a double
black hole binary. For details see the text.
 }
\label{bhbh}
\end{figure}

In considering the standard model of population synthesis (model
A) we evolved $N_{tot}=10^7$ initial binaries. All other models
were were run with $N_{tot}=4\times 10^6$ initial binaries. The
number of coalescing compact object binaries obtained in each
simulation is listed in Table~1. An example of an evolutionary
path leading to formation of a double black hole binary is
presented in Figure~\ref{bhbh}.  In this particular example,
the evolution starts with two
massive stars (primary mass $M_1=53.0 M_\odot$, secondary mass
$M_1=28.4 M_\odot$) at Zero Age Main Sequence on a rather wide
(semi major axis of orbit $a=176 R_\odot$) and eccentric orbit
(eccentricity $e=0.4$). More massive primary evolves first off
Main Sequence (MS) and starts evolution on a giant branch. Due
to a rapid radial expansion, it fills its Roche lobe
circularizing the orbit and initiates the first mass transfer
(MT)  episode (stage II). This transfer  is stable: rather small
orbital contraction and we assume non-conservative evolution:
primary looses its entire envelope, half of which is lost from
the system and only the remaining half is attached to the
companion. After MT episode we are left with a bare Helium core
of the primary and now the more massive rejuvenated MS secondary
(stage III). Primary, evolves fast and ends its life in a SN Ib
explosion, inducing a slight eccentricity on the binary and
widening the orbit (IV). As a result of the core collapse/SN
event a first BH is formed (V). Now, the secondary follows the
early path of its companion. It leaves MS and becomes a giant
star. Radial expansion leads to the second MT episode, which due
to the extreme mass ratio of two components (stage VI), is
dynamically unstable. The system goes through a Common Envelope
(CE) phase, with drastic shrinkage of the orbit, while almost
entire secondary envelope is ejected from the system. Small
fraction of the envelope is accreted onto the BH and the
secondary becomes a massive helium star (stage VII). Further
nuclear burning of elements in the interior of the helium star
brings it finally to a core collapse and SN Ib explosion (stage
VIII). The second BH is formed, and we end up with two massive
stellar BHs orbiting each other on a tight and eccentric orbit
(stage IX).

For each compact object binary we note the masses of individual 
objects $m_1$ and $m_2$,and the lifetimes  $t_{life}$  of the
system from its creation on the zero age main sequence to
formation of the  q compact object binary and  then  due to
the gravitational wave energy loss.   Within our code the
distinction between a neutron star and a black hole is based
solely on its mass. In our calculations we have been assuming a
maximum mass of a neutron star $M_{max}^{NS}=3\,M_\odot$, all
compact objects heavier than $M_{max}^{NS}$ are considered to be
black holes. Given that we can distinguish three different types
of binaries in our calculations: double neutron star binaries
(NSNS), black hole neutron star binaries (BHNS) and double black
hole binaries (BHBH). It appears that some properties of these
binaries are different.

\begin{figure}
\includegraphics[width=0.9\columnwidth]{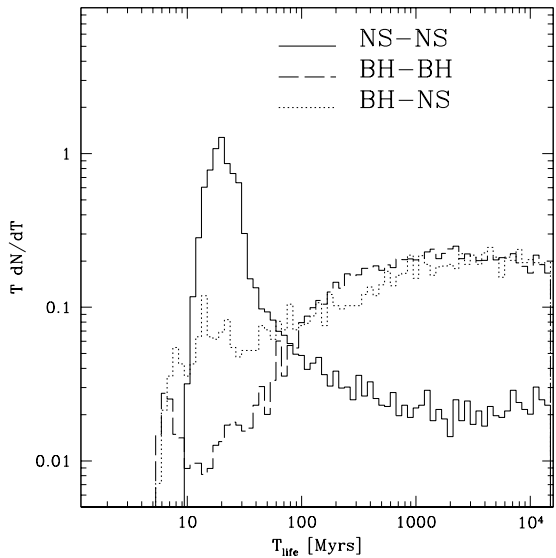}
\caption{ Distribution of the lifetimes of the double neutron star binaries
(solid line), black hole neutron star binaries (dotted line), and 
double black hole binaries (dashed lines).}
\label{times}
\end{figure}

We plot in Figure~\ref{times} the distribution of the lifetimes of these 
three types of binaries obtained within model A. 
 A majority
of  double neutron star binaries are rather short lived, with the lifetimes
in the range from $10$ to $50$\,Myrs \citep{2002ApJ...571L.147B}. 
The mixed BHNS binaries 
lifetimes span w very wide range from roughly $10$\,Myrs until the Hubble time
with no really preferred interval. The BHBH binaries live   
much longer their lifetimes extend from about $100$\,Myrs to the Hubble time.
Thus the currently merging NSNS binaries originate
in  stars formed in relatively recent starbursts, while
 the BHBH and BHNS binaries originate in stars that have been formed 
a few billion years ago! 
\citet{1997NewA....2...43L} and recently \citet{2003ApJ...589L..37B}
have shown that the observed sample of coalescing binaries is dominated by the
BHBH binaries.  
Since they are so long lived and the star formation rate 
 was probably much higher in the  early Universe  the accurate calculation 
 of their observed properties requires taking into account a full cosmological of
 distribution and evolution of such sources.

\begin{table*}
\caption{Description of different population synthesis models
used here. We list the number of binaries produced in each simulation.}

\begin{center}
\begin{tabular}{lp{7.7cm}r}
\hline \hline
Model & Description &   N produced\\
\hline
A      & standard model described in  \citet{2002ApJ...572..407B},
        but with $T_{Hubble}=15$Gyrs &33437 \\
B1     & zero kicks                  & 47690\\
B7      &  single Maxwellian with $\sigma=50$\,km\,s$^{-1}$  &47715\\
B11     & single Maxwellian with $\sigma=500$\,km\,s$^{-1}$ & 5246\\
B13    &   \citet{1990ApJ...348..485P}  kicks with $V_k=600$km\,s$^{-1}$  &5247\\
C      & no hyper--critical accretion onto NS/BH in CEs &  8416\\
E1     & $\alpha_{\rm CE}\times\lambda = 0.1 $  &  1003\\
E2     & $\alpha_{\rm CE}\times\lambda =   0.5 $ & 5538\\
E3     &$\alpha_{\rm CE}\times\lambda =   2$ &17595\\
F1     & mass fraction accreted: f$_{\rm a}=0.1 $ & 4461\\
F2     &  mass fraction accreted: f$_{\rm a}=  1$ &9885\\
G1     & wind changed by\ $f_{\rm wind}=0.5 $ & 12345\\
G2     & wind changed by\ $f_{\rm wind}=  2$ & 11482\\
J      & primary mass: $\propto M_1^{-2.35}$ & 13903\\
L1    & angular momentum of material lost in MT: $j=0.5 $  & 13529\\
L2    & angular momentum of material lost in MT: $j=  2.0$  & 3335\\
M1    & initial mass ratio distribution: $\Phi(q) \propto q^{-2.7} $ &  1363\\
M2    & initial mass ratio distribution: $\Phi(q) \propto   q^{3}$ & 22097\\
O      & partial fall back for $5.0 < M_{\rm CO} < 14.0 \,M_\odot$ & 9193\\
S      & all systems formed in circular orbits& 8306\\
Z1     & metallicity: $Z=0.01$ & 8520\\
Z2     & metallicity:   $Z=0.0001$ & 10481 \\

\end{tabular}

\end{center}
\end{table*}

\section{Expected numbers of observable mergers}

\begin{figure}
\includegraphics[width=0.9\columnwidth]{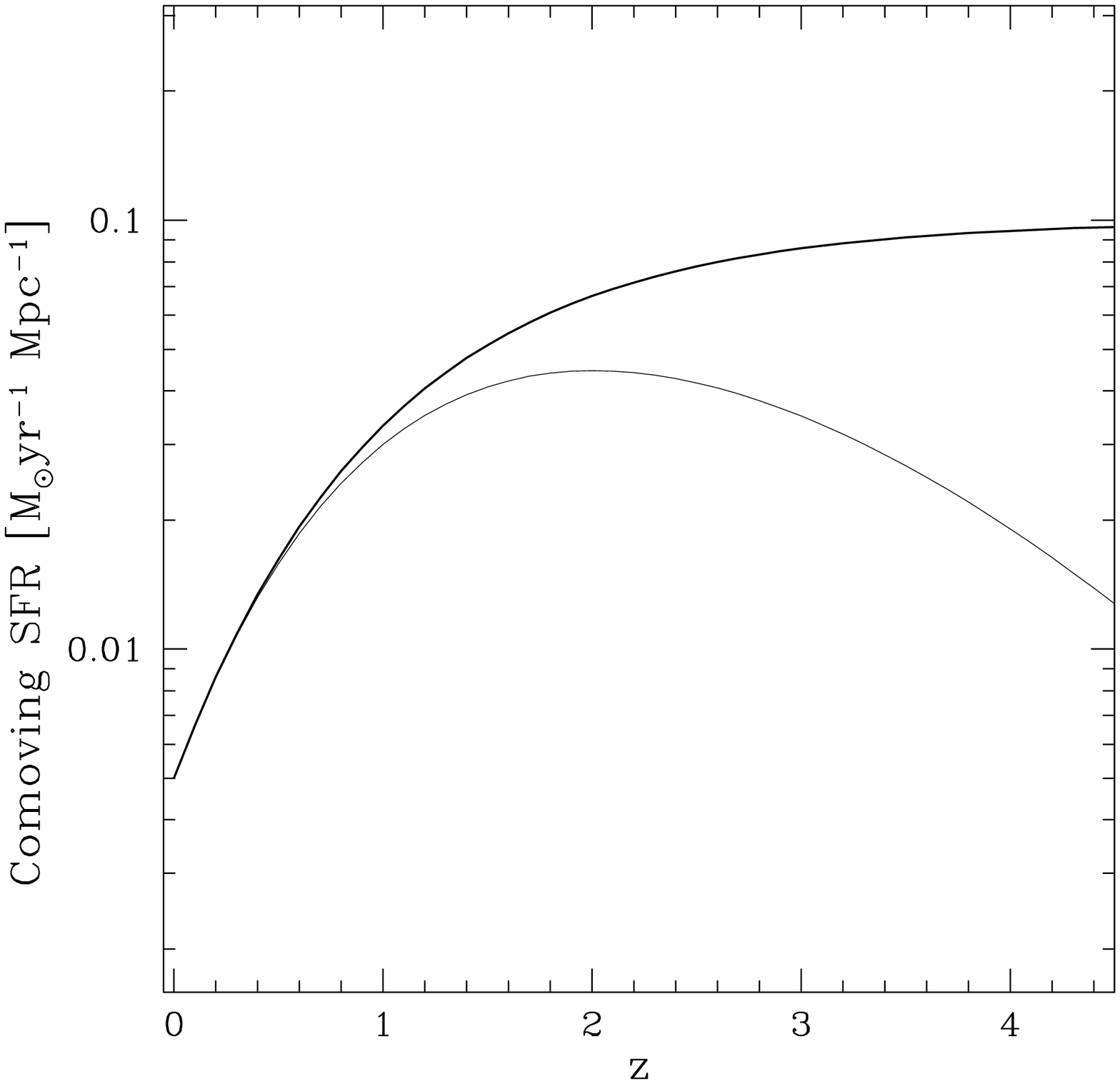}
\caption{The   star formation rate history models 
used here: thin line  \protect\citep{1996MNRAS.283.1388M}
, thick line \protect\citep{1999Ap&SS.266..291R}}
\label{sfr}

\end{figure}

The signal to noise from a coalescence of 
a binary in gravitational wave interferometers
has been calculated by several authors
\citep{1993ApJ...411L...5C,1994Bon,1998PhRvD..57.4535F}.
In the frequency region where the detectors are most sensitive 
 it is
 \begin{equation}
(S/N) = {A_i\over d_L } \left({(1+z) {\cal M}\over M_\odot}\right)^{5/6}\, ,
\label{sn}
\end{equation}
 where
${\cal M} = (m_1 m_2) ^{0.6} (m_1+m_2)^{-0.2}$ is the chirp mass,
$z$~is the redshift, $d_L$ is the luminosity distance, and the $A_i$
is the distance at which a coalescence of a $1\,M_odot$ binary
produces $(s/N)=1$ in the detector.
In a more detailed calculation - see \citet{1998PhRvD..57.4535F}
for the cases of LIGO I, LIGO II, and LISA -   equation~(\ref{sn}) 
becomes
more complex due to the finite size of the 
frequency interval available for a given detector.
This formula is valid with the accuracy better than 10 percent
for chirp masses below ${\cal M}< 20\, M_\odot$.
In the following  calculations we assume a flat  cosmological model
with the matter density $\Omega_m=0.3$, and cosmological constant
$\Omega_\Lambda=0.7$, the Hubble constant $H_0$ is
$65$\,km$^{-1}$Mpc$^{-1}$. We assume a model of star formation history rate 
presented as a thick solid line in Figure~\ref{sfr} \citep{1999Ap&SS.266..291R},
but we also consider a model in which the star formation rate ceases
at $z>2$ - thin line in Figure~\ref{sfr} \citep{1996MNRAS.283.1388M}, 
and a flat star formation rate.

Finding
the sampling redshift $z_{\cal M}$ requires solving  equation~(\ref{sn}),
with a required value of $(S/N)$,
where we insert $d_L=(1+z)d_{prop}$, and $d_{prop}$ is the proper distance
\begin{equation}
d_{prop}(z)=\int_0^z (1+z) c\left|{dt \over dz}\right| dz \, .
\end{equation}
Here we have introduced the
cosmic time  $t$:
\begin{equation}
\left|{dt \over dz}\right|=
{1\over  H_0 (1+z) [(\Omega_m(1+z)^3 +(1-\Omega_m)]^{1/2}}\, ,
\label{zt}
\end{equation}
in the flat space time.

Formation of the   compact object binaries is directly
connected with the star formation rate. After a binary is formed
it evolves for a time $t_{life}$ until its merger. 
This time includes the stellar evolutionary time needed to form
 a compact object
binary and then the life time of such binary due
to gravitational wave emission. The coalescence 
is delayed by  $t_{life}$ with respect to the 
star formation. Let us denote  by $F({\cal M},t,z)$
the formation rate of binaries with the chirp mass
 ${\cal M}$ and a
lifetime  $t$ as a function of redshift.
The  star formation rate history in the Universe
$SFR(z)$ can now be combined with the results
of the population synthesis code to find $F({\cal M},t,z)$.
To this end we need two quantities: the average stellar
mass $\langle M_*\rangle$, so that we find the
star number formation rate $SFR(z)/\langle M_* \rangle$, and the 
fraction of the stellar population that we simulate with the stellar
population synthesis code $f_{sim}$, to obtain 
the number formation rate of our binaries as a function of 
$z$:  $SFR(z) f_{sim}/\langle M_* \rangle$.
For the assumed slope of the initial mass
function   $\alpha=-2.7$, the minimal stellar mass of $0.2\,M_\odot$,
maximal stellar mass of $100\,M_\odot$, and binary fraction of $0.5$,
the average stellar mass is $\langle M_* \rangle =0.87\,M_\odot$, and
the fraction of stars that we simulate is
$f_{sim} =1.24\times 10^{-3}$.  The numerical 
estimate of the formation rate of binaries in a given interval
between  ${\cal M}$ and ${\cal M}+d {\cal M}$, with lifetimes
between $t$ and $t+dt$ is 
\begin{equation}
F({\cal M},t,z)d{\cal M} dt = SFR(z) {f_{sim}\over \langle M_* \rangle }
{N_i \over N_{tot}  }
\end{equation}
where $N_i$ is the number of binaries in a
 simulation  with the chirp mass and the lifetime in a given interval
and $N_{tot}$ is the total number of simulated binaries.

Binaries coalescing at a given redshift $z_0$   originate from binaries
formed at different earlier times.
The  rate of   coalescences of binaries
with a given chirp mass ${\cal M}$   is then given by
\begin{equation}
{d f_{coal} \over d{\cal M} }(z_0,{\cal M})
=\int dt' F({\cal M},t',z_f) \, ,
\end{equation}
where  the source
formation 
redshift $z_f$ is obtained by solving:
\begin{equation}
t' = \int_{z_0}^{z_f} \left| {dt \over dz}\right| dz \, . 
\end{equation}

We can now proceed to calculation of the   observed rate 
of coalescences.
This calculation is very similar to 
the ones performed in the case of gamma-ray bursts 
\citep{1999ApJ...511...41T,2002ApJ...571..394B}. An instrument
can detect signal characterized by the chirp mass ${\cal M}$
if it lies closer than the  sampling redshift $z_{\cal M}$
The rate at which such instrument will detect
binary coalescences is
\begin{equation}
{dR \over d{\cal M}} = \int_0^{z_{\cal M} }
{d f_{coal} \over d{\cal M} }(z,{\cal M}') {1\over 1+z} {dV\over dz}
dz
\label{rate}
\end{equation}
where ${\cal M}'= {\cal M}(1+z)^{-1}$ and
\begin{equation}
{dV\over dz}=4\pi c {d_L^2 \over (1+z)} \left|{dt \over dz}\right|
\end{equation}
is the comoving volume element.

\begin{figure}
\includegraphics[width=0.9\columnwidth]{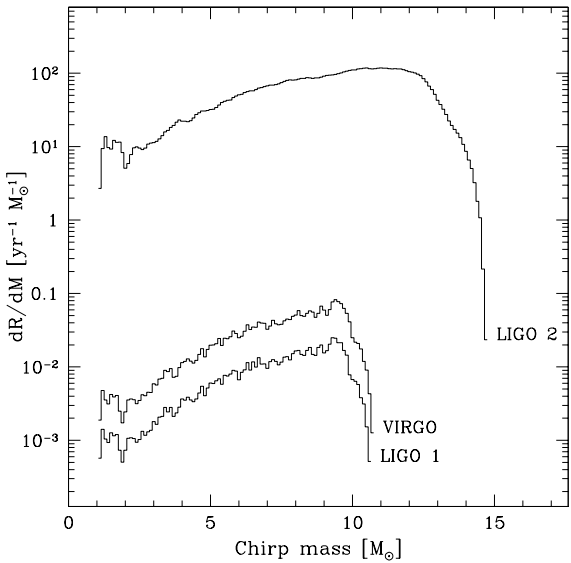}
\caption{The differential detection 
rates expected for the existing long base interferometers.}
\label{rates}
\end{figure}

 The results of the 
differential rate calculation are shown in Figure~\ref{rates}
where we present the model ${dR\over d{\cal M}}$ requiring a
signal to noise $(S/N)>8$ and  using the 
sensitivities calculated by \citet{1998PhRvD..57.4535F},
 LIGO I and LIGO II  and by \citet{phello98} for VIRGO, i.e.:
$A_{LIGO~1}=100$\,Mpc, $A_{VIRGO}=150$\,Mpc, and $A_{LIGO~2}=2200$\,Mpc.
There two main effects influencing the ${dR\over d{\cal M}}$
presented in Figure~\ref{rates}: the first  is that with increasing 
sensitivity a detector detects coalescences from a
 larger volume and the second is
due to the fact that the observed quantity is the redshifted chirp mass
$(1+z){\cal M}$. A highly sensitive detector like LIGO 2
shall detect coalescing binaries at non negligible redshifts and 
this leads to the spread of the distribution of observed chirp masses 
to higher values.

At this point we must also note the uncertainties
in the estimates shown in Figure~\ref{rates}, especially  
coming from the product $SFR(z) f_{sim}  \langle M _* \rangle^{-1}$.
The estimates of the star formation rates are uncertain by a factor
of at least two \citep{1996MNRAS.283.1388M}. Both
$f_{sim}$ and $\langle M_* \rangle$ depend strongly on the 
assumed slope of the initial mass function, and the lower mass cutoff of the
stellar population, and weakly on the binary fraction.
The low mass stars,
not taken into account in our simulations since they do 
not lead to compact object binaries, dominate the stellar
population in numbers and in mass.
We have assumed a power
law shape of the initial mass function, yet the absolute numbers
of the high mass progenitors of compact object binaries will be
affected by deviation of the initial mass function from a simple  power
law. A deviation of the slope by $1/2$ changes the number of 
high mass stars in relation to the total population by 
a factor of ten!
 Let us summarize: In the calculation of the rate 
using equation~(\ref{rate}), there are at least   the following
uncertainties: the unknown binary fraction, could be as high as a
factor of two, the unknown shape of the initial mass function
leads to uncertainty by a factor of up to ten. 
The differences between the galactic  
coalescence rates obtained in the framework
of various models of stellar evolution \citep{2002ApJ...572..407B}
amount to a factor of about $  30$. The difference in the expected rate
within various  stellar evolution models is comparable to
the intrinsic uncertainty in the rate. Thus given a measurement
of a rate, answering  an inverse problem, i.e. 
what does the rate tell us about the underlying stellar
evolution, shall be next to impossible. Here we have assumed that
the measured
rate has no uncertainty, yet we know already that the
interferometric detectors exhibit a very non stationary noise
which hampers an honest estimate of the time space volume 
surveyed.

\section{Distribution of chirp masses}

We note that the shape of inferred distributions
of chirp masses in Figure~\ref{rates} is nearly identical for
the LIGO 1 and VIRGO detectors. It is therefore interesting to consider
the distribution of the observed chirp masses as a potential
observational statistic. One obvious advantage of such
distribution is that all the problems  concerning the rate are
suddenly vanishing. The uncertainties in the normalization  that
entered the calculation of the rate vanish when the distribution
of  observed chirp masses is considered. We shall now consider
the sensitivity of the distribution of observed chirp masses to
the cosmological model assumed, and to the sensitivity of a
detector. Finally we will present the sensitivity of this
statistic to the  assumed model of stellar evolution.

We will start by defining two reference models with which other 
models will be compared.  These models corresponds to the cosmological model
described in section 2, with the star formation rate given
by the thick line in Figure~\ref{sfr}, and the sensitivity 
of a detector given by equation~\ref{sn}, 
with $A_1=100$\,Mpc for   model $\bf 1$ and $A_2=1$\,Gpc for model  $\bf 2$ .
In the following we will use
a simple method  of comparing distributions: the Kolmogorov Smirnov
test. This test uses a parameters $D$ defined
as the maximum distance between two cumulative
distributions. Two distributions differing by $D$ can be distinguished 
at a confidence level of approximately $10^{-4}$, when they are
sampled at $N\approx 4/D^2$ points. 

\begin{figure}
\includegraphics[width=0.9\columnwidth]{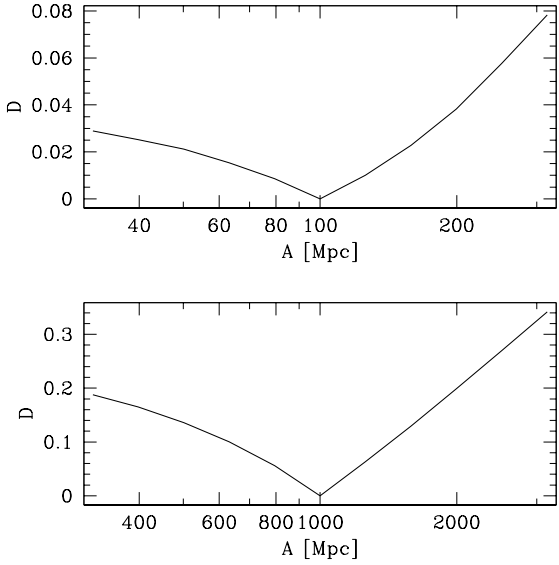}
\caption{Variation of the shape of the distribution of observed chirp masses
with the sensitivity of a detector. The top panel corresponds to the reference
model $\bf 1$, and the bottom panel
to the reference model $\bf 2$.
On the vertical axis we plot the Kolmogorov Smirnov parameter $D$ 
comparing a given distribution with the standard model described in the text.  }
\label{ddep}
\end{figure}

We first analyze the dependence of the 
expected distribution of observed chirp masses
on  detectors sensitivity. Variation in the
 sensitivity of a detector
is modeled by the parameter $A_i$ in equation~\ref{sn}.
We present the results in Figure~\ref{ddep} for our two reference models.
For the case of the reference model $\bf 1$ (top panel of Figure~\ref{ddep}),
the maximum value of the parameter $D$ reaches $\sim 0.04$ when
the sensitivity is increased or decreased by   a factor of two.
For the more sensitive detector - reference model $\bf 2$ in the bottom panel
of Figure~\ref{ddep} - the difference are larger and the parameter $D$ reaches
the value of $\sim 0.15$. This is due to the fact that the redshift effects play
a much stronger role for the more sensitive detectors. A smaller 
affects is connected with the fact that 
the detector starts seeing   NSNS mergers from redshifts where the star
formation rate was larger than locally.

\begin{figure}
\includegraphics[width=0.9\columnwidth]{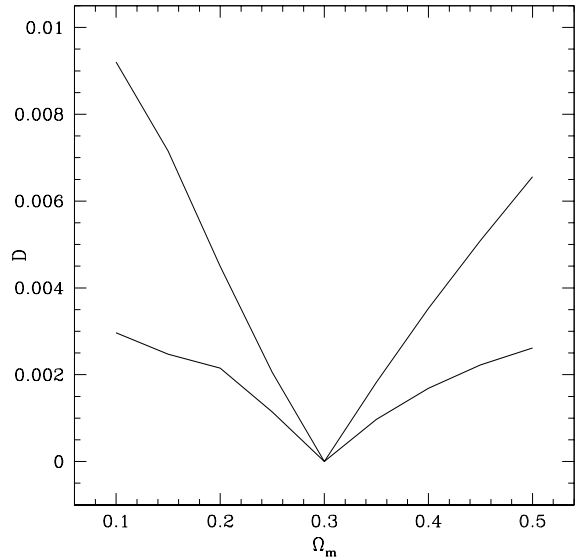}
\caption{Change in the distribution of the observed chirp masses
as a function of the assumed mass density in the cosmological model. 
The thick line corresponds to the reference model $\bf 1$, and the thin line to the
reference model $\bf 2$.}
\label{omdep}
\end{figure}

\begin{figure}
\includegraphics[width=0.9\columnwidth]{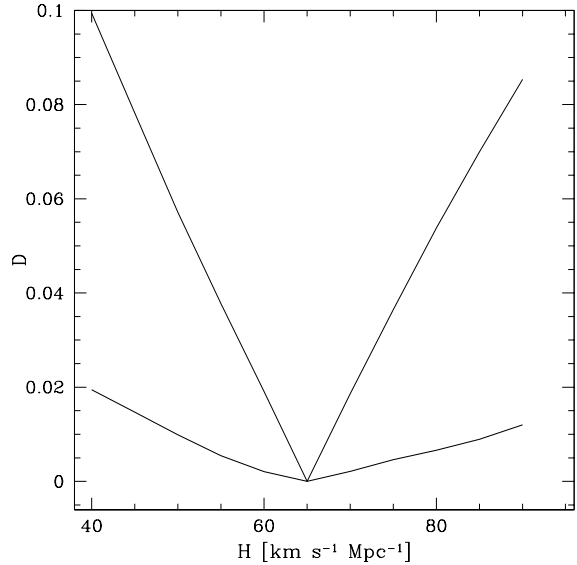}
\caption{Change in the distribution of the observed chirp masses
as a function of the assumed value of the Hubble constant.
The thick line corresponds to the reference model $\bf 1$, and the thin line to the
reference model $\bf 2$.
}
\label{hdep}
\end{figure}

We present the dependence of the shape of the
distribution of observed chirp masses on the assumed cosmological model
in  Figures~\ref{omdep} and~\ref{hdep}. The thick lines denote
the case the  reference  model $\bf 1$ with $A_1=100$\,Mpc, and
we present the reference model $\bf 2$
$A_2=1 $\,Gpc with thin lines. Varying $\Omega_m$ (while keeping
the Universe flat, i.e. $\Omega_\Lambda=1-\Omega_m$)
  does
not alter the shape of the observed distribution significantly:
the parameter $D$ never exceeds $0.01$ for model $\bf 2$
and less than $0.003$ for model $\bf 1$.
The assumed value of the Hubble constant has a stronger effect, however
when $H_0$ is varied by $20$\% the parameter $D$
does not exceed $0.01$ for the model $\bf 1$ and
$0.04$ for the model $\bf 2$.
We have also investigated the dependence of the
distribution of observed chirp masses on the
assumed shape of the star formation rate history.
Apart from the standard model, with the star formation rate history given by
the thick line in Figure~\ref{sfr}, we investigated a model where
the star formation rate
ceases above $z=2$, shown as a think line in Figure~\ref{sfr},
and a model where the star formation rate is constant.
The differences between the resulting distribution for
the  reference  model $\bf$  and the model with star formation decreasing above
$z=2$ leads to $D=0.011$, and for the flat star formation rate
we obtained
$D=0.021$, while for model $\bf 2$ these differences are: $D=0.010$ and $D=0.036$,
respectively.

\begin{figure}
\includegraphics[width=0.9\columnwidth]{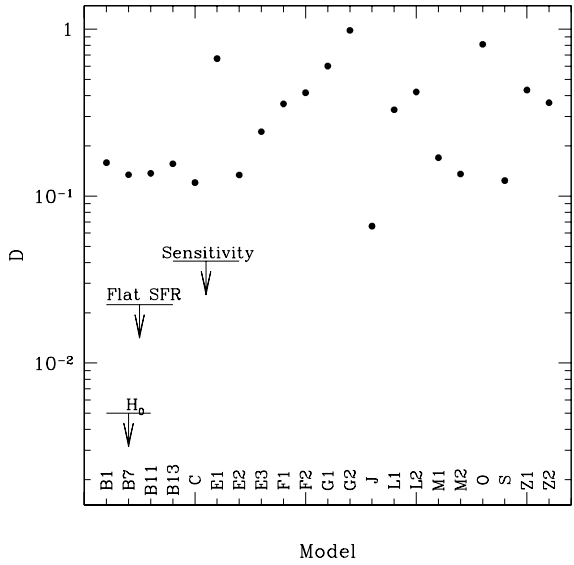}
\caption{The dependence  of the parameter $D$ comparing
the distribution of observed chirp masses of the reference model $\bf 1$
  with other models of stellar evolution.
For comparison we also present the maximal
values of $D$ obtained when cosmological parameters are changed,
for different star formation models, and when the sensitivity of the detector
varies.
 }
\label{mdepC}
\end{figure}

\begin{figure}
\includegraphics[width=0.9\columnwidth]{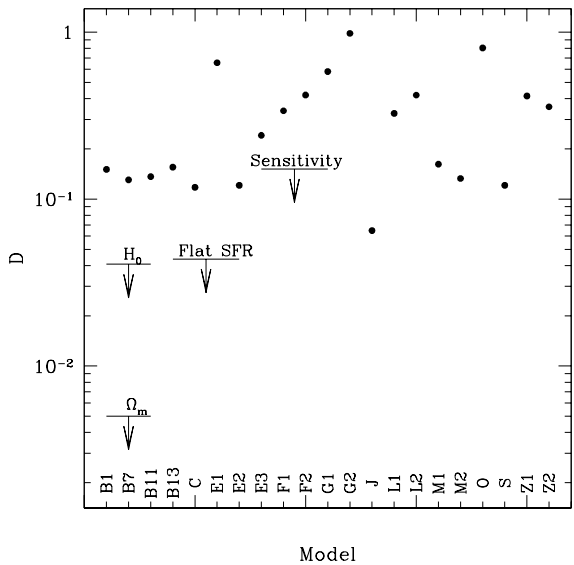}
\caption{Same as Figure~\protect{\ref{mdepC}} but for reference model $\bf 2$.
 }
\label{mdepM}
\end{figure}

We summarize the results of this section in Figures~\ref{mdepC}
and~\ref{mdepM}. Here we plot the values of the parameter $D$ obtained
from comparing our reference model $\bf 1$ (Figure~~\ref{mdepC}), and $\bf 2$
(Figure~~\ref{mdepM})
with distributions obtained when the stellar evolution is described by models
listed in Table~1.
 We also present the maximal values of the
parameter $D$ obtained when we varied  the detector sensitivity
is varied,    the cosmological model,
and also for different assumed forms of star
formation rate history.
  For most of the models the value
of $D$ lies between $0.1$ and unity. The exception is model J
for which $D=0.07$. This is the model with different slope of
the initial mass function.  The parameter $D$ is small in this
case since the distribution of observed chirp masses is
dominated by heavy double black hole mergers. Decreasing the
slope of the initial mass function only enhances this effect,
since it increases the ratio of black hole formation rate to
neutron star formation rate. For a number of models: E1, G1, G2,
and O the parameter  $D$ is  larger than $0.5$. These are the
models which vary from the standard one  by these parameters
that affect the masses of compact object the most. The remaining
models have the $D$ parameter in the range between $0.1$ and
$0.5$. Distributions differing by $D=0.2$  can be distinguished
when sampled with approximately one hundred  points. Thus we
confirm  our previous conclusion  \citep{2003ApJ...589L..37B}
that observation of about 100 mergers shall yield significant
constraints on the models of evolution of high mass stars.  In
Figure~\ref{mdepC} the expected distribution of chirp masses 
is mostly influenced by  the underlying 
stellar evolution model, while in the case of reference model two 
shown in Figure~\ref{mdepM}, variation of the sensitivity of the detector
may influence the shape of the chirp mass distribution at the level 
similar to several stellar models.
 Thus we
conclude that the distribution  of observed chirp masses is a
rather robust estimator of the underlying stellar evolution.

\section{Summary}

 Using the StarTrack  binary population
synthesis code we have investigated the properties of the population of double
compact objects - the primary candidate sources of gravitational
waves for high frequency interferometric detectors.  We find
that the distribution of lifetimes of various  types of
binaries  is different: double neutron star binaries live
typically for  a few tens of million years, while the lifetimes
of binaries containing  black holes extend up to the Hubble
time. Therefore, the double black hole binaries merging
currently in our Milky way neighborhood originate  in systems
formed even in the early star formation episodes! Hence, a
proper calculation of the number and properties of the  merging
compact object binaries should include the full cosmological
model of  formation of these sources. We outline such
calculation and calculate the expected  differential rate per unit
observed chirp mass. We discuss the  usefulness of the observed
rate for determining the  properties of the underlying
population. We estimate the  systematical
uncertainty of the calculation of the rates due to uncertainty
in the model. We find that the rate calculation carries a huge
systematic error, comparable to the spread in the rate  due to
variation of the stellar evolution model. We conclude that it is
very unlikely that significant constraints  can be obtained from
consideration and modeling of the rate alone. However, we
analyze the properties of another observable, the distribution
of observed chirp masses. We show that is very insensitive to
the parameters that made the rate estimate  uncertain. Moreover, we
test    the sensitivity of this statistic to the assumed
parameters of the  cosmological model: the mass density in the
Universe, the value of the Hubble constant, and the model of
star formation rate. All of these parameters hardly affect the
shape of the distribution of observed chirp masses. We also 
verify that the shape of  this distribution does not vary
significantly  when  a detectors sensitivity changes. This is
especially important for detectors with non stationary noise.
Non stationary noise would make comparison of the theoretical
rate with  observations even more uncertain and difficult, yet
it poses no problem for the analysis of the distribution of
observed chirp masses. Analysis of the distributions of the
observed chirp masses is therefore a very valuable  tool for
using the gravitational wave data to impose constraints on the
stellar evolution models.

The changes of the distribution of observed chirp masses with 
varying detectors sensitivity are mainly due to the fact that the observed
quantity is the redshifted chirp mass. Yet a redshift of a coalescing 
sources may be measured provided that we know its location
\citep{1993PhRvD..47.2198F}. A location can be estimated 
using a network of gravitational wave detectors. A measurement  of 
of the redshift would lead to the estimate of the chirp mass, and to removal of
the bulk of the dependence of the observed chirp mass distribution 
on the fluctuations of detectors sensitivity.

We must note that in general each measurement of a coalescence
may  also carry more information, like e.g. the individual
masses of the coalescing object or their spins. Once these
become available they must be included in the analysis. In this
paper we have considered only the chirp mass measurement, as
this is the most conservative approach.

\begin{acknowledgements}
This research was supported by the KBN grant 
5P03D01120.
\end{acknowledgements}


\end{document}